\documentclass{jcgt}

\setciteauthor{Stefan Zellmann and Jeff Amstutz}
\setcitetitle{A Practical Guide to Implementing Off-Axis Stereo Projection
Using Existing Ray Tracing Libraries}
\submitted{\today}

\accepted{submitted}{accepted}{published}{Editor Name}{vol}{issue}{1}{1}{year}
\seturl{http://jcgt.org/published/vol/issue/num/}

\usepackage[capitalize]{cleveref}
\usepackage{tikz}

\begin{document}

\title{A Practical Guide to Implementing Off-Axis Stereo Projection Using Existing
Ray Tracing Libraries}

\author
       {Stefan Zellmann\\University of Cologne
        \and Jefferson Amstutz\\NVIDIA
       }

\teaser{
  \includegraphics[width=2in]{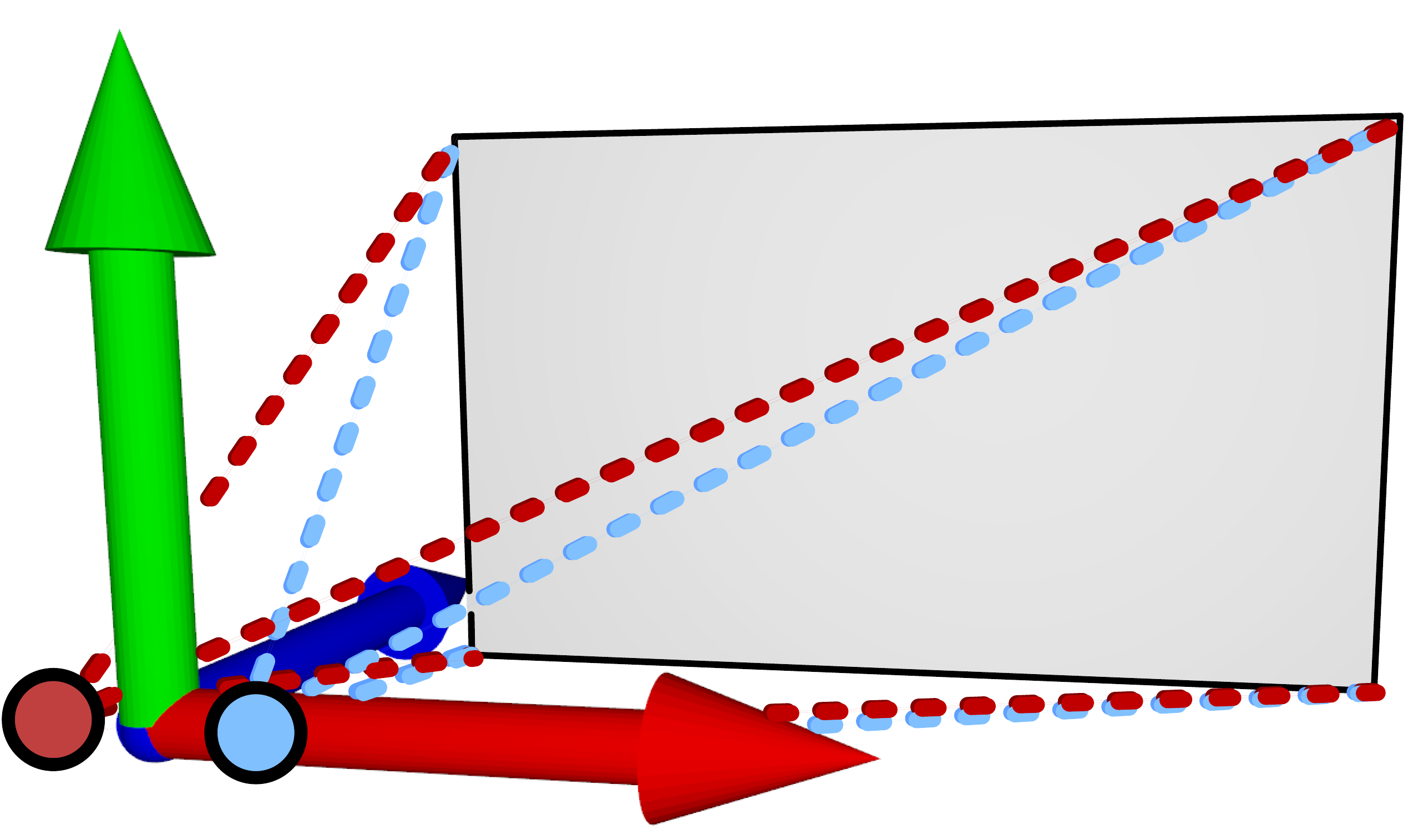}\hspace{1.2em}
  \includegraphics[width=2in]{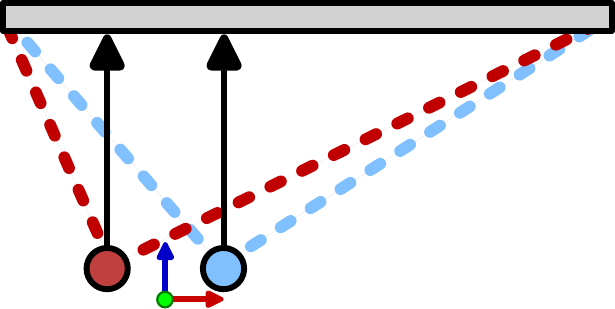}
  \caption{CAVE virtual environments use off-axis stereo projection so the user
  can move freely in front of the viewing plane. Off-axis projection is not
  supported by the perspective camera model that many ray tracing libraries
  use. We show how to still implement off-axis projection without having to
  adapt the internal implementation of the ray tracing library used.}
  \label{fig:teaser}
}

\maketitle
\thispagestyle{firstpagestyle}

\begin{abstract}
\small
%%This document is the \textit{JCGT} formatting template \textit{and} a description of the guidelines
%%for  \textit{JCGT} articles.  The author names on this document are fictional and chosen to show
%%how to format affiliations.
Virtual reality (VR) renderers driving CAVEs and similar immersive environments
use the off-axis stereo camera model so that a tracked user can move freely in
front of the projection plane. Geometrically, off-axis projection results in
asymmetric viewing frusta and generalizes the ubiquitous perspective camera
model to support positioning off the center of the projection plane. VR
renderers often integrate with larger visualization systems that rely on
libraries for position tracking and pose estimates, for ray tracing-based
rendering, and for user interaction. We demonstrate different strategies to
implement off-axis stereo projection within the constraints of given VR
applications and ray tracing libraries. We aim for minimal to no adjustments
required to the internal camera representation of such libraries. We include
host and shader code with the article that can be directly integrated in custom
applications.
\end{abstract}

%-------------------------------------------------------------------------
\section{Introduction}
\label{sec:introduction}
In this article we give a practical guide how to implement stereo projection
camera models within the constraints of given ray tracing libraries and virtual
reality applications. We assume an \emph{off-axis camera model} where the
projection planes (e.g., the walls of a CAVE virtual
environment~\cite{cruz-neira}) remain stationary and the camera position and
orientation change relative to that; the geometric setup can be fully described
by three corners of the rectangular projection plane and the camera position
and interpupillary distance of the eyes represented by the virtual camera in
world space.

How to set up this projection with OpenGL is described by
Pape~\cite{pape:2005}. In the article we follow the notation by Pape and
particularly refer to the world coordinate system defined by the projection
planes as \emph{CAVE space}. An implementation that is compatible with OpenGL
and glm that generates modelview and projection matrices on the host is given
in \cref{lst:gl}. Here, the input parameters are the lower-left (\texttt{LL}),
lower-right (\texttt{LR}), and upper-right (\texttt{UR}) corners of a fixed
screen, and the \texttt{eye} position, which is perpendicular and moves
relative to it. These parametesr can, e.g., be provided in metric units such as
meters or inches. The output matrices (\texttt{projOUT} and \texttt{viewOUT} in
the listing) define the coordinate transformation from CAVE space to OpenGL
window coordinates and can be used to transform objects in the shader code of
an existing rasterization pipeline.
\begin{lstlisting}[caption={Setting up off-axis stereo projection with OpenGL on the host (cf.\ \protect\cite{pape:2005}).}, label={lst:gl}, float]
void offaxisStereoTransform(vec3 LL, vec3 LR, vec3 UR, vec3 eye,
                            mat4 &projOUT, mat4 &viewOUT)
{
    vec3 X = (LL-LL)/length(LR-LL);
    vec3 Y = (UR-LR)/length(UR-LR);
    vec3 Z = cross(X,Y);

    // from world to eye coords (eye at origin, looking down -Z)
    mat3 R = inverse(mat3(X,Y,Z);

    // eye position relative to screen
    vec3 eyeP = eye-LL;

    // distance from eye to screen/wall
    float dist = dot(eyeP,Z);

    float left   = dot(eyeP,X);
    float right  = length(LR-LL)-left;
    float bottom = dot(eyeP,Y);
    float top    = length(UR-LR)-bottom;

    float znear = 1e-3f, zfar = 1000.f; // not relevant to us here

    left   = -left*znear/dist;
    right  =  right*znear/dist;
    bottom = -bottom*znear/dist;
    top    =  top*znear/dist;

    projOUT = frustum(left, right, bottom, top, znear, zfar);
    viewOUT = mat4(vec4(X,0.f),vec4(Y,0.f),
                   vec4(Z,0.f),vec4(-eye,1.f));
}
\end{lstlisting}

Existing and ubiquitous ray tracing libraries such as OSPRay~\cite{ospray} or
ANARI~\cite{anari}, even if they support stereo camera models, typically do not
allow for moving and orienting the camera relative to the projection
plane~\cite{ospray-immersive}. Instead they assume a centered camera, similar
to what was described by Suffern~\cite{suffern-stereo}. This model uses
off-axis projection internally, but does not expose the parameters to set up
the asymmetric frusta to the user. As such, the camera model is useful, e.g.,
for head-mounted diplays (HMD), but not for CAVE, powerwall, or powerbench
virtual environments used in engineering or other scientific applications.

Another potential burden is that some existing virtual reality applications
do not provide direct access to the off-axis camera parameters, yet they expose
plug-in mechanisms to extend the application with custom logic. Virtual
reality renderers targeting CAVE and similar virtual environments are often
written against rasterization APIs~\cite{paraview,cover,osg}. In that case, the
plug-ins might be allowed to implement viewport extensions, including custom
renderers that write directly to the frame buffer, but have only indirect
access to the viewing transform (in the form of OpenGL matrices) inside the
render loop or shader programs.

Our article provides a practical guide to extending existing ray tracing APIs
and virtual reality renderers that do not expose dedicated off-axis camera
models to their plug-ins to support CAVE-style stereo rendering, and whose
internal source code is unavailable or cannot be changed. We provide host and
device code along with the article that can directly be copied into plug-ins
and as such is of use to practitioners who integrate commodity ray tracing
tools into existing virtual reality applications.

\section{Three Off-Axis Stereo Camera Implementation Strategies}
\label{sec:algos}
In this section we describe three different strategies to implement off-axis
stereo projection, and the associated set of primary rays to be used in a ray
tracer, given different combinations of the constraints described above in
\cref{sec:introduction}.

\subsection{Strategy 1: Inverse Viewing Transform to Primary Rays}
\label{sec:strategy1}
The first strategy to compute primary rays using off-axis projection aims at
solving the case where the camera is given by two matrices---the projection
matrix with intrinsic and the model/view matrix with extrinsic camera
parameters---as well as by the viewport transform, inside the render loop or
shader program. This strategey is briefly mentioned in Zellmann's
thesis~\cite{zellmann:phd}.

The premise behind the approach is that an arbitrary camera model---including
but not limited to off-axis---is encoded by the transforms. The transforms are
used to convert given 3D objects (including points and vectors) from world
space to window coordinates using OpenGL; i.e., just as one could transform 3D
surfaces, it is also possible to transform a set of primary rays from world to
window coordinates, and hence, given the inverse transform, it is then also
possible to transform the rays \emph{back} to world (and also: CAVE) space.
\begin{lstlisting}[caption={Shader code transforming orthogonal rays in OpenGL
normalized device coordinates using the inverse viewing transform, to give
primary rays in CAVE space.}, label={lst:matrices}, float]
Ray primaryRayFromViewingTransform(
    int x, int y, mat4 projInv, viewInv, int width, int height)
{
    float u = 2 * (x+0.5f) / width - 1.f;
    float v = 2 * (y+0.5f) / height - 1.f;

    vec4 o = viewInv * (projInv * vec4(u, v, -1, 1));
    vec4 d = viewInv * (projInv * vec4(u, v,  1, 1));

    Ray r;
    r.ori = o.xyz / o.w;
    r.dir = normalize(d.xyz / d.w - r.ori);
    return r;
}
\end{lstlisting}

The strategy exploits this by generating orthogonal viewing rays in the
\emph{normalized device} unit coordinate system (NDC). $u/v$-positions in the
horizontal and vertical dimensions are obtained by normalizing the viewport's
$x/y$ pixel positions to the range $[-1,1]$; the rays originate at $z=-1$ and
point towards $z=1$. Shader sample code is given in \cref{lst:matrices}. A
real-world application might set up the projection and viewing matrices with
code similar to \cref{lst:gl}, and a plug-in might, e.g., obtain the matrices
using a call to \texttt{glGet()} inside the render loop where the OpenGL
context is active, invert them, and pass them on to the shader where the
primary rays are generated.

We assume that the depth range associated with the viewport transform is
$[0,1]$ and does not require further handling by the ray generation function.
Although the near and far clipping plane do not influence depth precision as
they would in a rasterizer, their values will manifest in the projection
matrix; our model can handle any reasonable user-provided values as long as all
the geometry is visible (if the user desires to really use these planes for
clipping, we recommend to compute world-space intersections with the clip
planes by setting up \texttt{r.tmin} and \texttt{r.tmax} accordingly).

\subsection{Strategy 2: Converting Off-Axis to Stereo Pinhole Camera Pairs}
\label{sec:strategy2}
Off-axis projection is an extension to the ubiquitous perspective or pinhole
camera model. This model is used to implement stereo rendering by positioning
two cameras at eye positions in space that are given by the main camera (or,
head) position, and their interpupillary distance. Rendering the image for one
eye, off-axis generalizes the perspective camera model by positioning the
camera origin freely with respect to the viewing plane. Since the viewing
direction is still perpendicular to the viewing plane (an arbitray orientation
is achieved by positioning the left and right eye positions at an angle), the
frusta generated by this model are in general asymmetric (cf.\
\cref{fig:teaser}).

Suffern~\cite{suffern-stereo} proposes to implement this via a generalized
perspective camera model that supports translating the camera position relative to
the given image plane. The camera models by common ray tracing libraries such
as OSPRay~\cite{ospray} or ANARI~\cite{anari} unfortunately do not allow this.
Those libraries however support defining rectangular sub-image regions that
rendering is restricted to; in ANARI, for example, this is exposed as an
additional parameter called ``image region'' that defines a sub-image in the
range $[0,1]^2$. The ANARI devices (including OSPRay) implement sub-image
regions by interpolating and stretching the pixel contribution out on the image
plane when a sub-image region of size $<1$ is set, via the linear interpolation
operation
\begin{equation}
\begin{pmatrix} u \\ v\end{pmatrix} = \vec{1}-\begin{pmatrix} u \\ v\end{pmatrix}
\odot I_{lo} + \begin{pmatrix} u \\ v\end{pmatrix} \odot I_{hi},
\end{equation}
where $u,v$ are normalized pixel positions, $I_{lo}$ and $I_{hi}$ are the lower
and upper coordinates of the sub-image region, and $\odot$ is a component-wise
vector multiplication. This transformation is applied right before the primary
rays are calculated, based on $u,v$, and if the image region is unspecified,
$u,v$ remain unchanged after the transformation.

\begin{figure}[htb]
  \centering
   \includegraphics[width=0.5\columnwidth]{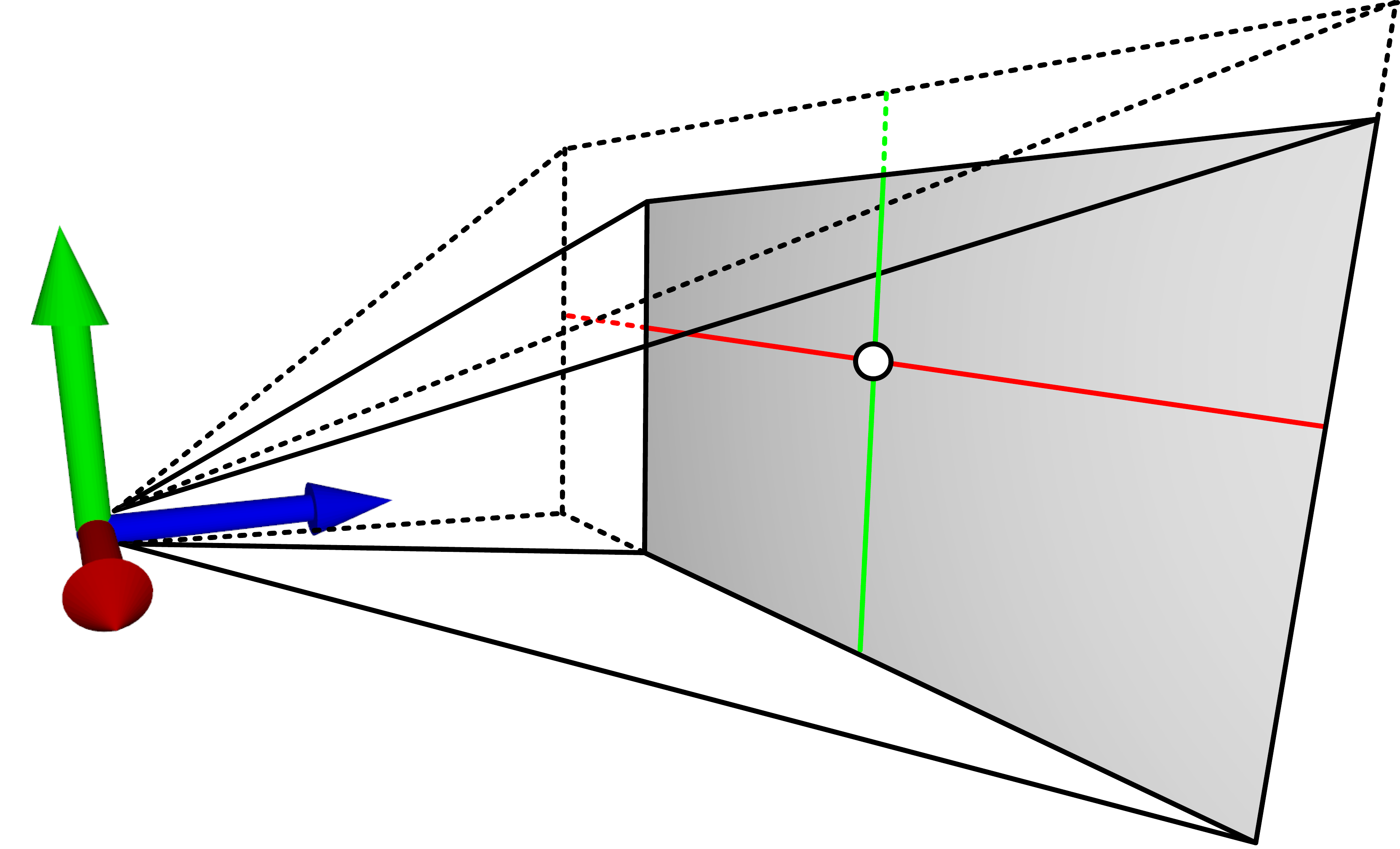}
   \caption{\label{fig:offaxis-with-sub-images}
The camera position projects to a window coordinate off the image center (black
circle), resulting in an asymmetric viewing frustum indicated by the solid
black lines. To render this view using an ordinary perspective camera model, we
increase the aspect ratio and field of view so that the frustum becomes
symmetric (dashed lines), and set up a sub-image region that only includes the
shaded region of the image corresponding to the visible screen.
}
\end{figure}
Using this sub-image region feature, it is possible to compute an off-axis
projection even if the perspective camera model does not support arbitrary
translation of the viewing position relative to the image plane. This can be
achieved by setting a wider aspect ratio and field of view, but at the same
time clipping the output image using sub-image regions. The geometrical setup
for this is shown in \cref{fig:offaxis-with-sub-images}, for the case where the
eye position is offset to the left of the image center, and at a height
slightly above the center \footnote{The case that the eye position projects to
a viewing plane coordinate outside the viewport of the application is not
considered here, as the positional tracking systems of our target hardware
cover about the same range as the viewing frusta.}. The analogous cases with
offsets to the right or bottom follow from symmetry.
\begin{lstlisting}[caption={Function that, given three corners of the viewing
plane in CAVE space, and the eye position, calculates the extrinsic
(\texttt{dir} and \texttt{up}) as well as intrinsic (\texttt{fovy} and
\texttt{aspect}) camera parameters, as well as a clip region to make
the resulting frustum asymmetric again.}, label={lst:strategy2}, float]
void offaxisStereoCamera(vec3 LL, vec3 LR, vec3 UR, vec3 eye,
                         vec3 &dirOUT, vec3 &upOUT,
                         float &fovyOUT, float &aspectOUT,
                         box2f &imageRegionOUT)
{
    vec3 X = (LR-LL)/length(LR-LL);
    vec3 Y = (UR-LR)/length(UR-LR);
    vec3 Z = cross(X,Y);

    dirOUT = -Z;
    upOUT = Y;

    // eye position relative to screen/wall
    vec3 eyeP = eye-LL;

    // distance from eye to screen/wall
    float dist = dot(eyeP,Z);

    float left   = dot(eyeP,X);
    float right  = length(LR-LL)-left;
    float bottom = dot(eyeP,Y);
    float top    = length(UR-LR)-bottom;

    float newWidth = left<right ? 2*right : 2*left;
    float newHeight = bottom<top ? 2*top : 2*bottom;

    fovyOUT = 2*atan(newHeight/(2*dist));

    aspectOUT = newWidth/newHeight;

    imageRegionOUT.min.x = left<right ? (right-left)/newWidth : 0.f;
    imageRegionOUT.max.x = right<left ? (left+right)/newWidth : 1.f;
    imageRegionOUT.min.y = bottom<top ? (top-bottom)/newHeight: 0.f;
    imageRegionOUT.max.y = top<bottom ? (bottom+top)/newHeight : 1.f;
}
\end{lstlisting}

\cref{lst:strategy2} presents an exemplary implementation to set up the
perspective camera given the same lower-left~(\texttt{LL}),
lower-right~(\texttt{LR}), and upper-right~(\texttt{UR}) corners as well as the
\texttt{eye} position from \cref{lst:gl} as input, and that outputs the
remaining extrinsic and intrinsic camera parameters, as well as a sub-image
region. 

Analogous to \cref{lst:gl} we first compute an orthonormal basis given by
\texttt{X}, \texttt{Y}, and \texttt{Z} that is centered at the lower-left
corner, with the z-axis pointing towards the viewer. These basis vectors give
us the remaining extrinsic parameters (direction and up vectors). Again
following \cref{lst:gl} we compute the distances from the camera position to
the viewing plane \texttt{dist}, and the distance of the projection to the
edges \texttt{left}, \texttt{right}, \texttt{bottom}, and \texttt{top}. These
allow us to compute the new field of view and aspect ratio by comparing which of
the distances is smaller; e.g., if the projected center is closer to the left
edge, the new, virtual width of the viewing plane is twice the distance to the
right edge (and similarly for the remaining cases). To compute the and
sub-image clip region we convert from CAVE space to unit distances by dividing
by the new viewport size.

With this strategy it is possible to integrate ray tracing libraries without
support for off-axis camera projection into existing VR applications, given
that these provide access to their viewing plane and off-axis camera
representations.

\subsection{Strategy 3: Generating Pinhole Cameras from Viewing Transforms}
\label{sec:strategy3}
The third strategy addresses the case where the ray tracing library exposes a
pinhole camera model only, and the application's camera parameters are only
available in the form of OpenGL transforms. This one is the most general of the
three strategies and can serve as a fallback in any given case. Code for this
strategy is provided in \cref{lst:strategy3}.
\begin{lstlisting}[caption={Function that computes the fixed screen corners and
eye position from give OpenGL transforms and feeds the result into the function
from \cref{lst:strategy2}.}, label={lst:strategy3}, float]
void offaxisStereoCameraFromXfm(mat4 projInv, mat4 viewInv,
                                vec3 &dirOUT, vec3 &upOUT,
                                float &fovyOUT, float &aspectOUT,
                                box2f &imageRegionOUT)
{
    // Transform NDC unit cube corners to world/CAVE space
    vec3 v000 = unprojectNDC(projInv, viewInv, vec3(-1,-1,-1));
    vec3 v001 = unprojectNDC(projInv, viewInv, vec3(-1,-1, 1));
    (...)
    vec3 v011 = unprojectNDC(projInv, viewInv, vec3(-1, 1, 1));
  
    // edges from +z to -z
    vec3 ez00 = normalize(v001-v000);
    vec3 ez10 = normalize(v101-v100);
    vec3 ez11 = normalize(v111-v110);
    vec3 ez01 = normalize(v011-v010);
  
    // edges from -y to +y
    (...)
  
    // edges from -x to +x
    (...)
  
    vec3 nL = normalize(cross(ey00, ez00));
    vec3 nR = normalize(cross(ez10, ey10));
    vec3 nB = normalize(cross(ez00, ex00));
    vec3 nT = normalize(cross(ex10, ez01));
  
    // Line of intersection between left/right planes
    vec3 pLR, nLR;
    intersectPlanePlane(nL, v000, nR, v100, nLR, pLR);
  
    // Line of intersection between bottom/top planes
    vec3 pBT, nBT;
    intersectPlanePlane(nB, v000, nT, v010, nBT, pBT);
  
    // Line segment connecting the two intersecting lines
    vec3 p1, p2;
    closestLineSegmentBetweenTwoLines(nLR, pLR, nBT, pBT, p1, p2);
  
    vec3 LL = unprojectNDC(projInv, viewInv, vec3(-1,-1, 1));
    vec3 LR = unprojectNDC(projInv, viewInv, vec3( 1,-1, 1));
    vec3 UR = unprojectNDC(projInv, viewInv, vec3( 1, 1, 1));
    vec3 eye = (p1+p2)/2.f;
  
    offaxisStereoCamera(LL, LR, UR, eye,
        dirOUT, upOUT, fovyOUT, aspectOUT, imageRegionOUT);
}
\end{lstlisting}

\begin{figure}[htb]
  \centering
  \begin{tikzpicture}
   \node[anchor=south west,inner sep=0] (image) at (0,0) {\includegraphics[width=0.99\columnwidth]{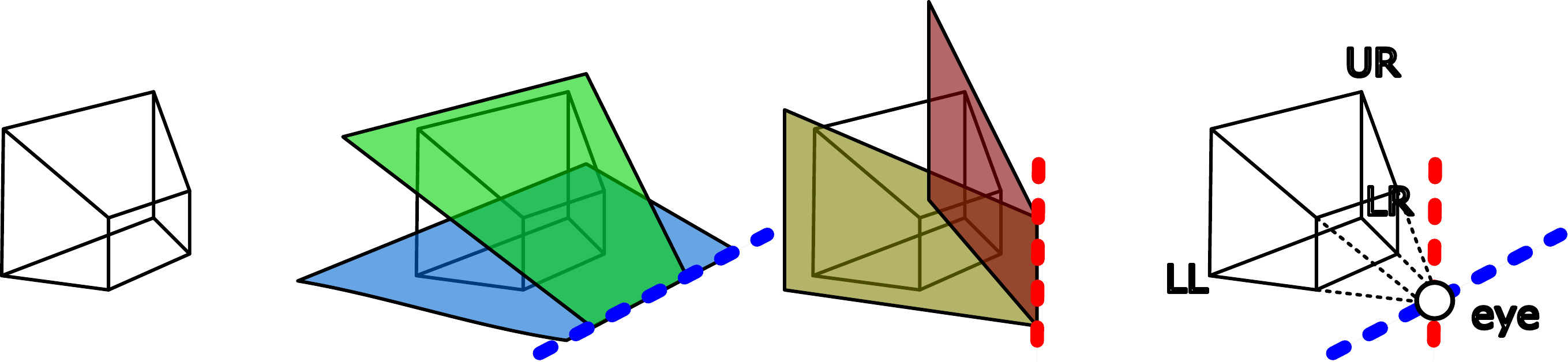}};
   \node[anchor=south west,inner sep=0] at (1.2,-0.6) {(a)};
   \node[anchor=south west,inner sep=0] at (4.5,-0.6) {(b)};
   \node[anchor=south west,inner sep=0] at (7.7,-0.6) {(c)};
   \node[anchor=south west,inner sep=0] at (11.0,-0.6) {(d)};
  \end{tikzpicture}
   \caption{\label{fig:strategy3}
Computing fixed-screen coordinates in CAVE space from given viewing transform.
(a)~Transforming the unit cube in NDC using OpenGL transforms gives us the
viewing frustum in CAVE space. We compute the intersecting lines of the
bottom/top~(b) and the \mbox{left/right}~(c) planes. The (conservative)
intersection of those intersecting line gives us the eye coordinate, which
together with the three corners defines the viewing configuration in CAVE
space~(d), and can be converted to a pinhole camera using the method from
\cref{sec:strategy2}.
}
\end{figure}

Here we make the assumption that the view to clip space transforms form a
perspective frustum, i.e., neither of the constituting planes are orthogonal.
This is the case for the types of cameras discussed so far. The geometric setup
of this strategy is illustrated in \cref{fig:strategy3}. We first define the
corners of the unit cube in NDC and transform them to CAVE space
(\cref{fig:strategy3}a). We use the transformed corners to compute the
\mbox{left/right} (\cref{fig:strategy3}b) and bottom/top (\cref{fig:strategy3}c)
planes of the asymmetric viewing frustum. We compute the intersecting lines
between the left/right and bottom/top planes, respectively, using the three
plane intersection algorithm~\cite{goldman-three-planes}. The intersecting
lines have a crossing at the eye position. To avoid numerical precision issues,
instead of using the intersection of the two lines in space, we conservatively
compute the shortest connecting line segment between the two plane
intersections and use the median of the two endpoints (which ideally are the
same) as the camera position.

The screen corners are given by the far plane ($z=1$) of the transformed unit
cube; note that generally, these are not the exact screen corners, but those of
a parallel rectangle with the same aspect ratio. Given the eye position and
three corners (\cref{fig:strategy3}d), we can compute the remaining extrinsic
and intrinsic pinhole camera parameters using \cref{lst:strategy2}.

\section{Evaluation}
\label{sec:eval}
We evaluate the three different strategies qualitatively, by integrating them
with different ray tracing libraries and rendering apps. We omit a performance
evaluation that we believe would not be meaningful in either way. In terms of
shader or device code, the second and first strategy share the exact same
implementation of generating primary rays using a pinhole camera; the performance
comparison would hence boil down to comparing pinhole camera ray generation to
ray generation via short vector math matrix multiplication. Besides, any
reasonable ray tracing application will not nearly be bottlenecked on ray
generation.

We base our evaluation on ANARI~\cite{anari} and extend different applications
to use it as a ray tracing rendering interface. Our implementations are tested
with the anari-ospray device~\cite{anari-ospray} and with VisRTX~\cite{visrtx}.
We further wrote an own prototypical ANARI device using the Visionaray
library~\cite{visionaray} that does not implement all the features, but allows
for more experimental code and for adding non-standard extensions
\footnote{https://github.com/szellmann/anari-visionaray}.

We provide an open source demo app with sample code
\footnote{\url{https://github.com/szellmann/anari-offaxis-sample-code}} that generates 3D
stereo views with fixed screen configurations and that uses ANARI. Here we
implemented the three camera models from \cref{sec:algos}. We can switch
between the different configurations; the only ANARI device supporting the
camera model from \cref{sec:strategy1} using the inverse viewing transform and
matrix multiplications is Visionaray; here we added a ``matrix'' camera
extension that we deem useful as a general camera extension to ANARI, since it
supports not only the perspective camera models described in the article, but
any camera model described by OpenGL, including orthographic projection. The
extension to ANARI comprises the API calls shown in
\cref{lst:anari-matrix-cam}. The API adds a new camera of type ``matrix'', and
allows to set the model/view and projection matrices obtained from OpenGL;
internally, this camera model stores inverted copies of these. Apart from the
Visionaray device, neither of the ANARI implementations supports the ``matrix''
camera extension, which limits portability of this solution.
\begin{lstlisting}[caption={ANARI extensions required to support
the ``matrix'' camera model from \cref{sec:strategy2}.}, label={lst:anari-matrix-cam}, float]
// Create camera of type "matrix"
ANARICamera cam = anariNewCamera(device, "matrix");

// Set the model/view and projection matrices
anariSetParameter(device, cam, "view", ANARI_FLOAT32_MAT4, &view[0]);
anariSetParameter(device, cam, "proj", ANARI_FLOAT32_MAT4, &proj[0]);
\end{lstlisting}

We extend the virtual reality renderer OpenCOVER~\cite{cover} with an ANARI
plug-in that can render fixed screen stereo camera pairs. OpenCOVER provides a
plug-in API with a class that provides the OpenGL transforms required to go from
CAVE space to window space and can be queried per eye and projection plane, in
the form of $4 \times 4$ matrices. The plug-in API also provides direct access
to the frame buffer for each eye and wall, either as a pointer on the CPU, or
as a CUDA ``device pointer''. Our plug-in consumes the camera transforms and
generates off-axis compatible pinhole camera pairs using the method from
\cref{sec:strategy3}.
\begin{figure}[htb]
  \centering
   \includegraphics[width=0.36\columnwidth]{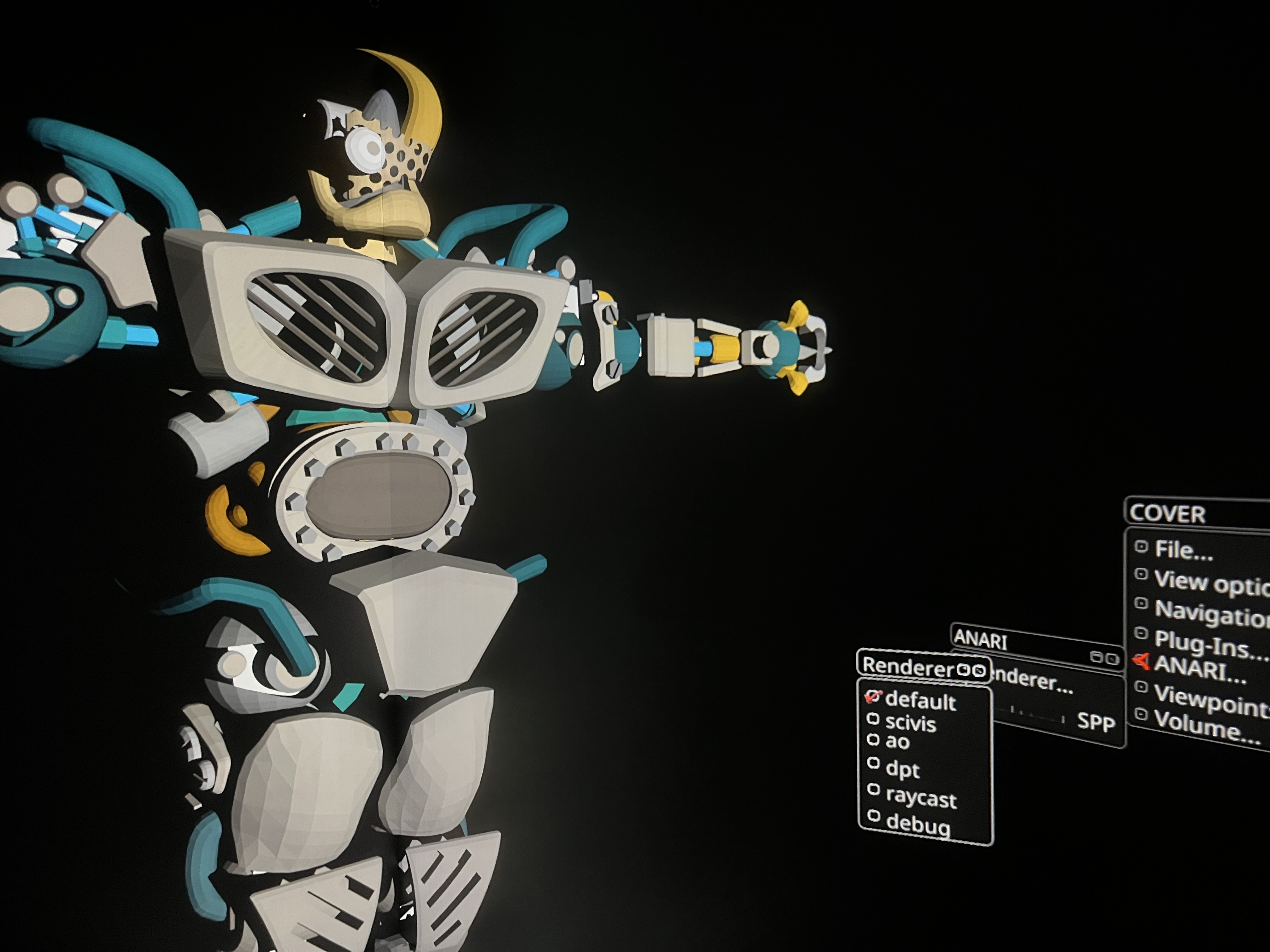}
   \includegraphics[width=0.36\columnwidth]{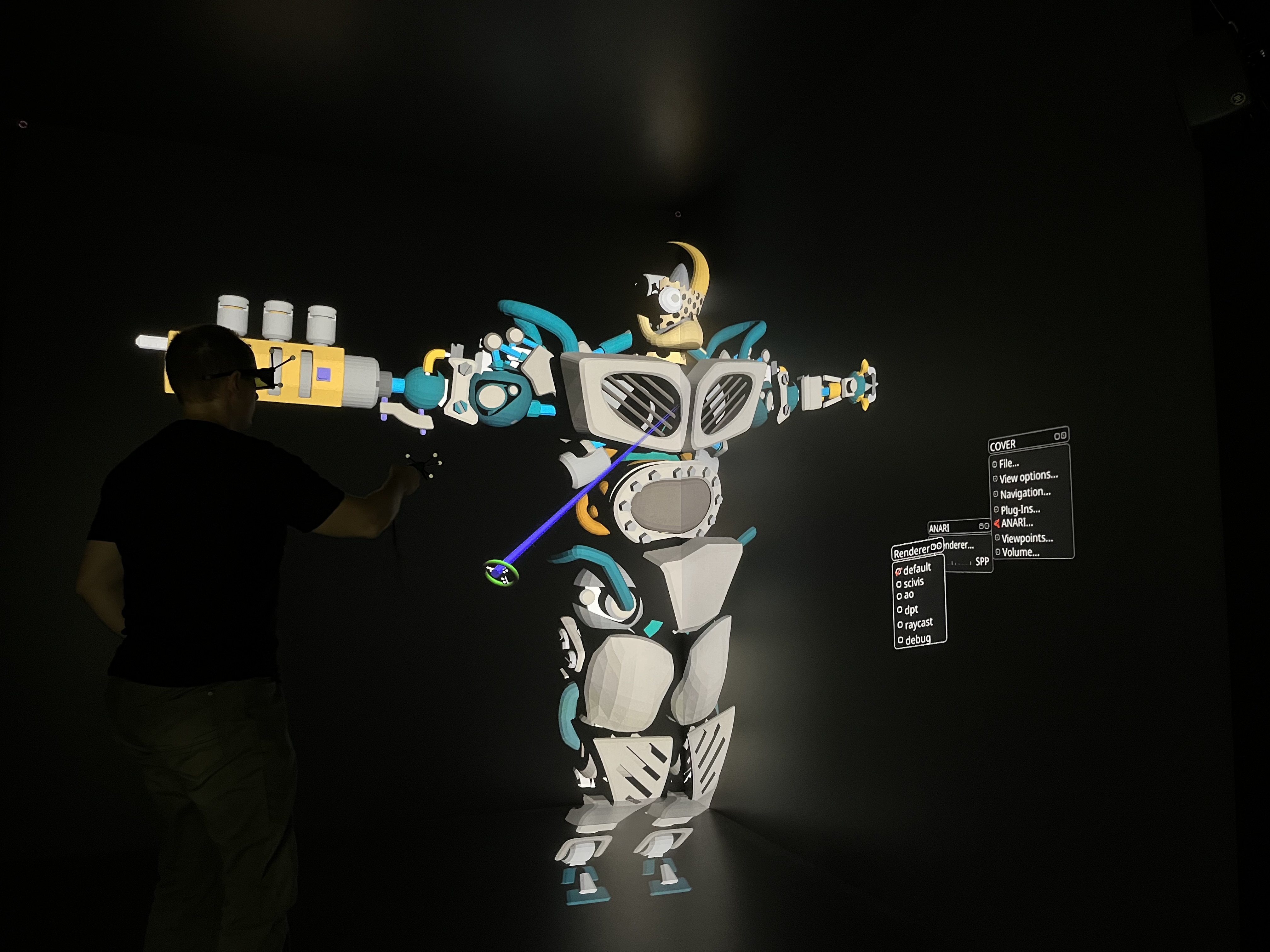}
   \includegraphics[width=0.2026\columnwidth]{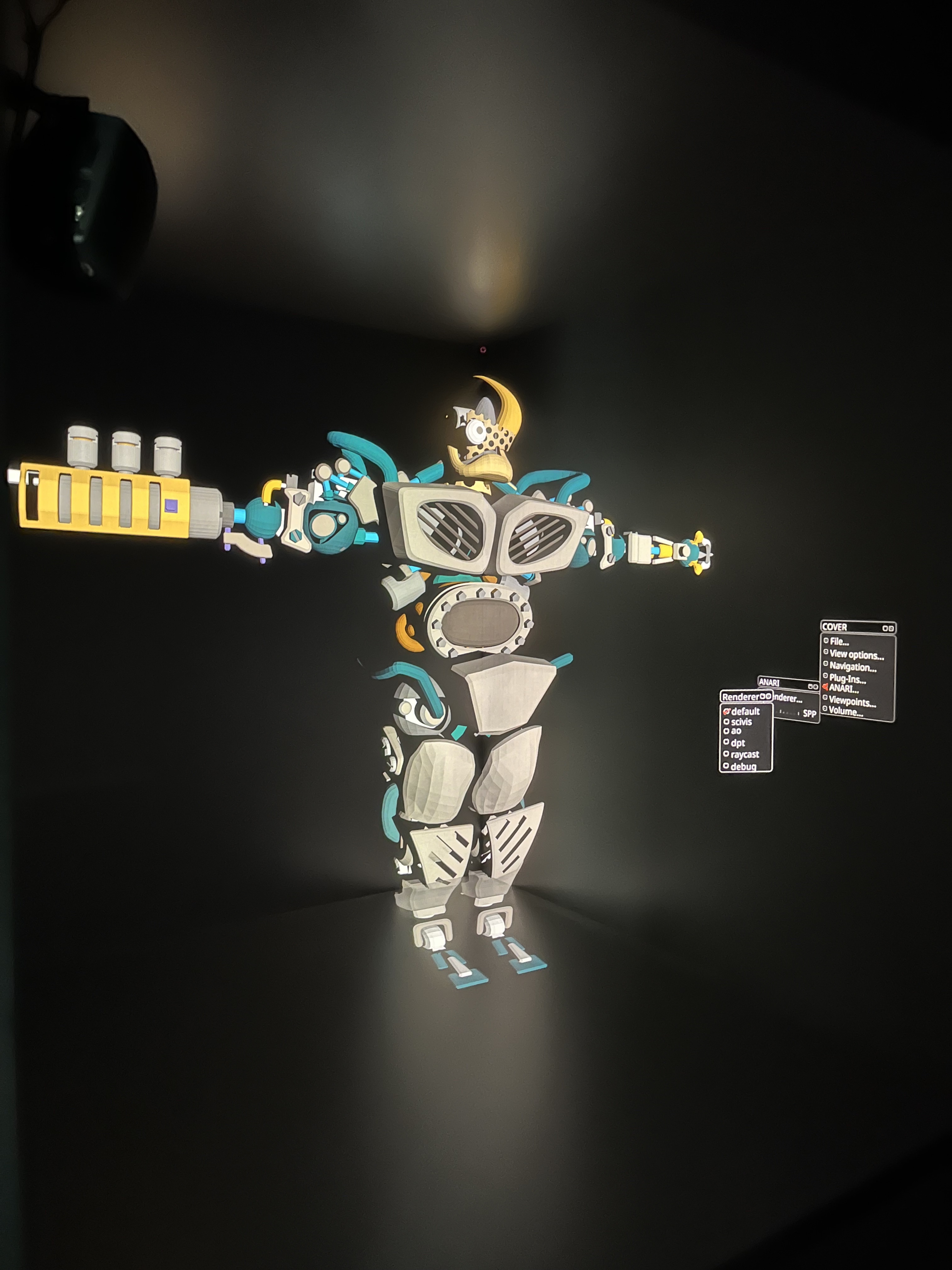}
   \caption{\label{fig:cave}
   ANARI/OpenCOVER integration of our method, running in the five-sided CAVE
   virtual environment at the University of Cologne.
}
\end{figure}
\cref{fig:cave} shows the plug-in in use rendering with VisRTX in the
five-sided CAVE virtual environment at the University of Cologne.  The hardware
configuration there uses active stereo projectors that are delivered
side-by-side stereo images of $1600 \times 1600$ pixels per eye per projection
plane. The system is operated by a cluster with eleven NVIDIA RTX~6000 GPUs
(Turing generation) (one GPU per eye, and a redundant view for the desk
operator).

\subsection{Limitations}
The three strategies to generate stereo camera pairs are general in that they
accommodate any viewing setup and ray tracing library, provided that the former
at least give access to the OpenGL viewing transform, and the latter implement
sub-image regions. We also found the performance of the three strategies
to be equivalent---we did not encounter performance regressions,
or numerical precision issues with either of the models, including the one
geometrically computing the screen corner and eye coordinates in CAVE space. A
limitation is that existing stereo camera implementations---which would also
facilitate compositing of stereo views---cannot be used; both
ANARI's~\cite{anari} and OSPRay's~\cite{ospray} pinhole cameras have a stereo
mode (which in the case of ANARI is however not implemented by a majority of
the existing devices). A limitation of the methods making use of the internal
pinhole camera model is that only perspective cameras are supported, while the
matrix camera extension would also support other projection types.

\section{Conclusion}
\label{sec:conclusion}
We presented three different strategies to implement off-axis stereo cameras to
generate primary rays for ray tracing in VR, with portability ranging from
barely to highly portable. Each strategy has its uses within the context of
different applications. We have validated their usefulness in the context of
existing ray tracing libraries and VR applications. The source code accompanying
the article is provided online and under a permissive open source license.

\section*{Acknowledgements}
This work was supported by the Deutsche Forschungsgemeinschaft (DFG, German
Research Foundation) under grant no.~456842964.

%All entries in the bibliography \textbf{must} have URLs unless no URL is possible (e.g., book, very old thesis).

\small
\bibliographystyle{jcgt}
\bibliography{paper}

\begin{thebibliography}{\protect\citename{Cruz-Neira et~al\mbox{.} }1992}

\bibitem[\protect\citename{Amstutz }2023]{visrtx}
{\sc Amstutz, J.}, 2023.
\newblock {VisRTX}.
\newblock Available at \url{https://github.com/NVIDIA/VisRTX}, Accessed: 21 Aug
  2023.

\bibitem[\protect\citename{Cruz-Neira et~al\mbox{.} }1992]{cruz-neira}
{\sc Cruz-Neira, C., Sandin, D.~J., DeFanti, T.~A., Kenyon, R.~V., and Hart,
  J.~C.}
\newblock 1992.
\newblock {The} {CAVE}: {Audio} {Visual} {Experience} {Automatic} {Virtual}
  {Environment}.
\newblock {\em Commun. ACM 35}, 6 (jun), 64–72.
\newblock URL: \url{https://doi.org/10.1145/129888.129892},
  doi:10.1145/129888.129892.

\bibitem[\protect\citename{Goldman }1990]{goldman-three-planes}
{\sc Goldman, R.}
\newblock 1990.
\newblock {\em {Graphics} {Gems}}.
\newblock Academic Press Professional, Inc., USA, ch.~{Intersection} of {Three}
  {Planes}, 305.

\bibitem[\protect\citename{Intel }2023]{anari-ospray}
{\sc Intel}, 2023.
\newblock {ANARI-OSPRay}---{Translation} {layer} from {Khronos} {ANARI} to
  {Intel} {OSPRay}: {ANARILibrary} and {ANARIDevice} ``ospray''.).
\newblock Available at \url{https://github.com/ospray/anari-ospray}, Accessed:
  21 Aug 2023.

\bibitem[\protect\citename{Martin et~al\mbox{.} }2018]{paraview}
{\sc Martin, K., DeMarle, D., Jhhaveri, S., and Ayachit, U.}, 2018.
\newblock {Taking} {ParaView} into {Virtual} {Reality}.
\newblock Available at
  \url{https://www.kitware.com/taking-paraview-into-virtual-reality/},
  Accessed: 13 August 2023.

\bibitem[\protect\citename{Nam et~al\mbox{.} }2023]{ospray-immersive}
{\sc Nam, J.~W., Abram, G.~D., Samsel, F., and Navrátil, P.~A.}
\newblock 2023.
\newblock {Immersive} {OSPRay}: {Enabling} {VR} {Experiences} with {OSPRay}.
\newblock In {\em PEARC: Practice and Experience in Advanced Research Computing
  (PEARC 2023)}.
\newblock preprint.
\newblock URL:
  \url{https://jungwhonam.github.io/images/publications/PEARC2023_Immersive-OSPRay.pdf},
  doi:10.1145/3569951.3597579.

\bibitem[\protect\citename{Pape }2005]{pape:2005}
{\sc Pape, D.}, 2005.
\newblock {Computing the CAVE Projection Transformation}.
\newblock Available at \url{https://www.evl.uic.edu/pape/caveproj/}, Accessed:
  10 Aug 2023.

\bibitem[\protect\citename{Rantzau and Lang }1998]{cover}
{\sc Rantzau, D., and Lang, U.}
\newblock 1998.
\newblock {A} {Scalable} {Virtual} {Environment} for {Large} {Scale}
  {Scientific} {Data} {Analysis}.
\newblock {\em Future Generation Computer Systems 14}, 3, 215--222.
\newblock Virtual Reality in Industy and Research.
\newblock URL:
  \url{https://www.sciencedirect.com/science/article/pii/S0167739X98000259},
  doi:https://doi.org/10.1016/S0167-739X(98)00025-9.

\bibitem[\protect\citename{Stone et~al\mbox{.} }2022]{anari}
{\sc Stone, J.~E., Griffin, K.~S., Amstutz, J., DeMarle, D.~E., Sherman, W.~R.,
  and Günther, J.}
\newblock 2022.
\newblock {ANARI}: {A} {3-D} {Rendering} {API} {Standard}.
\newblock {\em Computing in Science \& Engineering 24}, 2, 7--18.
\newblock URL: \url{https://ieeexplore.ieee.org/document/9745399},
  doi:10.1109/MCSE.2022.3163151.

\bibitem[\protect\citename{Suffern }2007]{suffern-stereo}
{\sc Suffern, K.}
\newblock 2007.
\newblock {\em {Ray} {Tracing} {from} {the} {Ground} {Up}}, 1st~ed.
\newblock A K Peters / CRC Press, ch.~12: Stereoscopy, 197--215.
\newblock URL:
  \url{https://www.taylorfrancis.com/chapters/mono/10.1201/b10675-12/stereoscopy-kevin-suffern}.

\bibitem[\protect\citename{Wald et~al\mbox{.} }2017]{ospray}
{\sc Wald, I., Johnson, G., Amstutz, J., Brownlee, C., Knoll, A., Jeffers, J.,
  Günther, J., and Navratil, P.}
\newblock 2017.
\newblock {OSPRay} - {A} {CPU} {Ray} {Tracing} {Framework} for {Scientific}
  {Visualization}.
\newblock {\em IEEE Transactions on Visualization and Computer Graphics 23}, 1,
  931--940.
\newblock URL: \url{https://ieeexplore.ieee.org/abstract/document/7539599},
  doi:10.1109/TVCG.2016.2599041.

\bibitem[\protect\citename{Yuan et~al\mbox{.} }2007]{osg}
{\sc Yuan, P., Wang, S., Zhang, J., and Liu, H.}
\newblock 2007.
\newblock {Virtual} {Reality} {Platform} {Based} on {Open} {Sourced} {Graphics}
  {Toolkit} {OpenSceneGraph}.
\newblock In {\em 2007 10th IEEE International Conference on Computer-Aided
  Design and Computer Graphics}, 361--364.
\newblock URL: \url{https://ieeexplore.ieee.org/abstract/document/4407909},
  doi:10.1109/CADCG.2007.4407909.

\bibitem[\protect\citename{Zellmann et~al\mbox{.} }2017]{visionaray}
{\sc Zellmann, S., Wickeroth, D., and Lang, U.}
\newblock 2017.
\newblock Visionaray: A cross-platform ray tracing template library.
\newblock In {\em 2017 IEEE 10th Workshop on Software Engineering and
  Architectures for Realtime Interactive Systems (SEARIS)}, 1--8.
\newblock URL: \url{https://ieeexplore.ieee.org/document/9183547},
  doi:10.1109/SEARIS41720.2017.9183547.

\bibitem[\protect\citename{Zellmann }2014]{zellmann:phd}
{\sc Zellmann, S.}
\newblock 2014.
\newblock {\em {Interactive} {High} {Performance} {Volume} {Rendering}}.
\newblock PhD thesis, University of Cologne, Albertus-Magnus-Platz, 50923
  Cologne, GER.
\newblock Chapter 5.1.1, pp. 81--86.
\newblock URL: \url{http://kups.ub.uni-koeln.de/5727/}.

\end{thebibliography}

\end{document}